\documentclass[lettersize,journal]{IEEEtran}

\usepackage{formatting/packages}
\usepackage{formatting/macros}

\begin{document}
\captionsetup{skip=0pt}
\setlength{\abovecaptionskip}{1pt}   
\setlength{\belowcaptionskip}{1pt}   
\setlength{\textfloatsep}{1pt}      
\setcounter{topnumber}{5}      

\title{\LARGE{\sysname: Accelerating Image Diffusion Transformer Inference with Mixed-Precision MX Quantization}}

\author{Daeun Kim, Jinwoo Hwang, Changhun Oh, \IEEEmembership{Member, IEEE}, Jongse Park, \IEEEmembership{Senior Member, IEEE}

\vspace{-0.2cm}
\thanks{Manuscript received February 27, 2025; revised April 7, 2025. This work was supported by Institute of Information \& communications Technology Planning \& Evaluation (IITP) (No.2018-0-00503, 33\%), ITRC(Information Technology Research Center) grant funded by the Korea government(MSIT) (IITP-2025-RS-2020-II201795, 33\%), and National Research Foundation of Korea(NRF) grant funded by the Korea government(MSIT) (RS-2024-00342148, 33\%). \textit{(Corresponding author: Jongse Park.)}
}
\thanks{The authors are with the School of Computing, Graduate School of AI Semiconductor, Korea Advanced Institute of Science and Technology, Daejeon 34141, South Korea (e-mail: dekim@casys.kaist.ac.kr; jwhwang@casys.kaist.ac.kr; choh@casys.kaist.ac.kr; jspark@casys.kaist.ac.kr)}
\thanks{Digital Object Identifier XXXXX.XXXXXXX}
\vspace{-1cm}
}

\markboth{IEEE Computer Architecture Letters, VOL. 24, No. 1, JANUARY-JUNE 2025}
{Shell \MakeLowercase{\textit{et al.}}: A Sample Article Using IEEEtran.cls for IEEE Journals}

\IEEEpubid{
  \begin{minipage}{\textwidth}
    \centering
    0000-0000~\copyright~2025 IEEE. Personal use is permitted, but republication/redistribution requires IEEE permission.\\
    See https://www.ieee.org/publications/rights/index.html for more information.
  \end{minipage}
}

\maketitle
\begin{abstract}
\underline{Di}ffusion \underline{T}ransformer (DiT) has driven significant progress in image generation tasks.
However, DiT inferencing is notoriously compute-intensive and incurs long latency even on datacenter-scale GPUs, primarily due to its iterative nature and heavy reliance on GEMM operations inherent to its encoder-based structure.
To address the challenge, prior work has explored quantization, but achieving low-precision quantization for DiT inferencing with both high accuracy and substantial speedup remains an open problem.
To this end, this paper proposes \sysname, an algorithm-hardware co-designed acceleration solution that exploits mixed Microscaling (MX) formats to quantize DiT activation values. 
\sysname quantizes the DiT activation tensors by selectively applying higher precision to magnitude-based outliers, which produce mixed-precision GEMM operations.
To achieve tangible speedup from the mixed-precision arithmetic, we design a \sysname accelerator that enables precision-flexible multiplications and efficient MX precision conversions.
Our experimental results show that \sysname delivers a speedup of 2.10–5.32$\times$ over RTX 3090, with no loss in FID.
\end{abstract}

\vspace{-0.2cm}
\begin{IEEEkeywords}
Diffusion transformer (DiT), Image generation, Quantization, Microscaling (MX), Acceleration
\end{IEEEkeywords}

\IEEEpubidadjcol
\vspace{-0.5cm}
\section{Introduction}
\begin{figure}[t]
\centering
\includegraphics[width=0.9\linewidth]{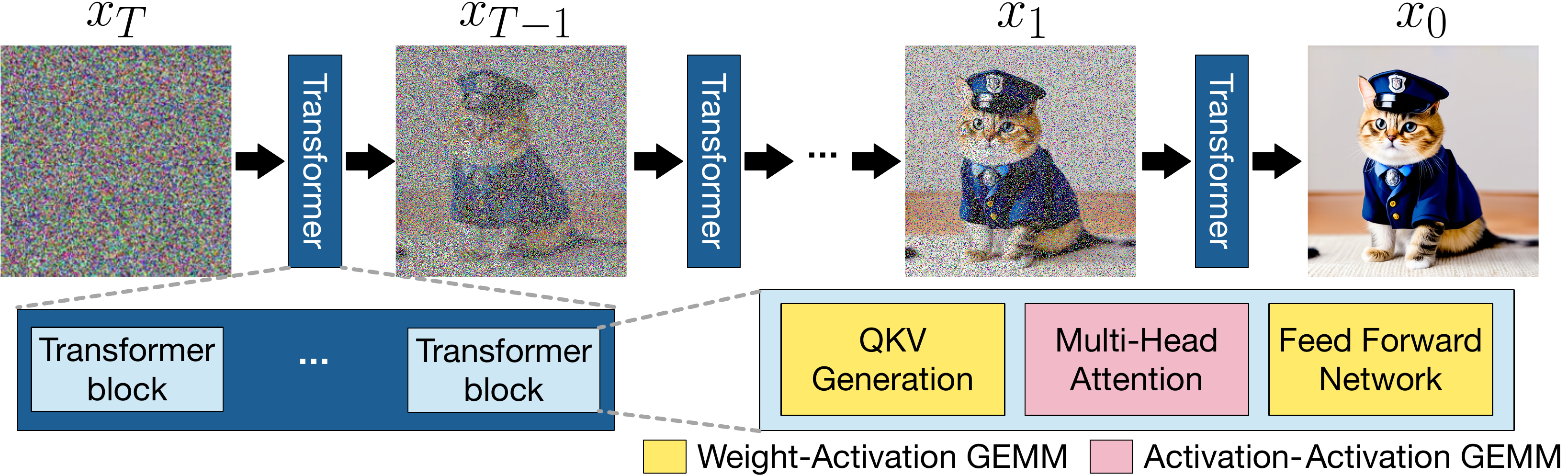}
\caption{Denoising process of image generation diffusion with total timestep $T$ and architecture of diffusion transformer.}
\label{fig:denoising}
\end{figure}

\vspace{-0.1cm}
%
\IEEEPARstart{R}{ecent} breakthroughs in diffusion models have sparked a paradigm shift in image synthesis by enabling more stable and high-fidelity generation. 
A key contributor to this shift is the integration of transformers into diffusion frameworks~\cite{sd3, pixart-sigma, flux}, also known as \underline{\textbf{Di}}ffusion \underline{\textbf{T}}ransformers (\emph{DiT}), building upon their proven success in both language and vision tasks.
Figure~\ref{fig:denoising} illustrates the denoising process of diffusion transformer model.
Random noise is iteratively and progressively denoised, ultimately resulting in a clear image. 
This denoising process necessitates repetitive transformer computations, which result in a notably slow inference process.
To address this challenge, several recent studies~\cite{ptq4dit, qdit, viditq, flightvgm, mixdq, paro} have explored quantization to accelerate diffusion models.
However, studies for DiT~\cite{ptq4dit, qdit, viditq} have struggled to enable quantization for \emph{activations}.

%
Rather disjointly, microscaling formats (MX)~\cite{mx, ocp} are emerging as a promising approach for quantization.
With its accuracy and hardware efficiency, MX has been standardized by Open Compute Project~\cite{ocp}, led by industry giants, and supported on NVIDIA Blackwell GPUs~\cite{blackwell}.
Although MX has shown effectiveness across various workloads, such as CNNs and LLMs, it remains unexplored in diffusion transformers.
In this study, we propose \sysname, an algorithm-hardware co-designed solution that leverages MX for low-precision quantization in diffusion transformers and introduces a specialized hardware architecture to support mixed-precision computation efficiently.
Our contributions are as follows: (1) magnitude-based mixed-precision MX quantization for aggressively low precision while preserving generation quality loss, (2) hyperparameter determination algorithm for mixed-precision thresholding, and (3) precision-flexible MX-supporting accelerator architecture. 
Our experiments showcase that \sysname achieves a latency speedup of 2.10$\times$ to 5.32$\times$ compared to NVIDIA RTX 3090, with no quality degradation in FID.

\IEEEpubidadjcol
\vspace{-0.6cm}
\section{Background and Motivation}
\subsection{Image Diffusion Transformer}
DiTs employ an encoder-based architecture, comprising Query-Key-Value (QKV) generation, multi-head attention layers, and feedforward networks (FFN). 
Iterative inferences of these layers incur high latency, even on datacenter-level GPUs.
For instance, we observe that Stable Diffusion 3 takes 18.75 seconds for generating a 1,024$\times$1,024 image on RTX-3090.

\vspace{-0.4cm}
\subsection{Quantization for Improving Performance}
\vspace{-0.1cm}
\niparagraph{Quantization for DiT.}
Quantization is a technique that converts floating-point representations into low-precision formats, which helps reduce the memory footprint and improve the computation efficiency.
Several prior works~\cite{ptq4dit, qdit, viditq} have explored the use of quantization for DiTs.
These studies typically use integer (INT) quantization, applying 4-bit for weights, while retaining 8-bit for activations, which limits speedup.

\begin{figure}[b]
\centering
\includegraphics[width=\linewidth]{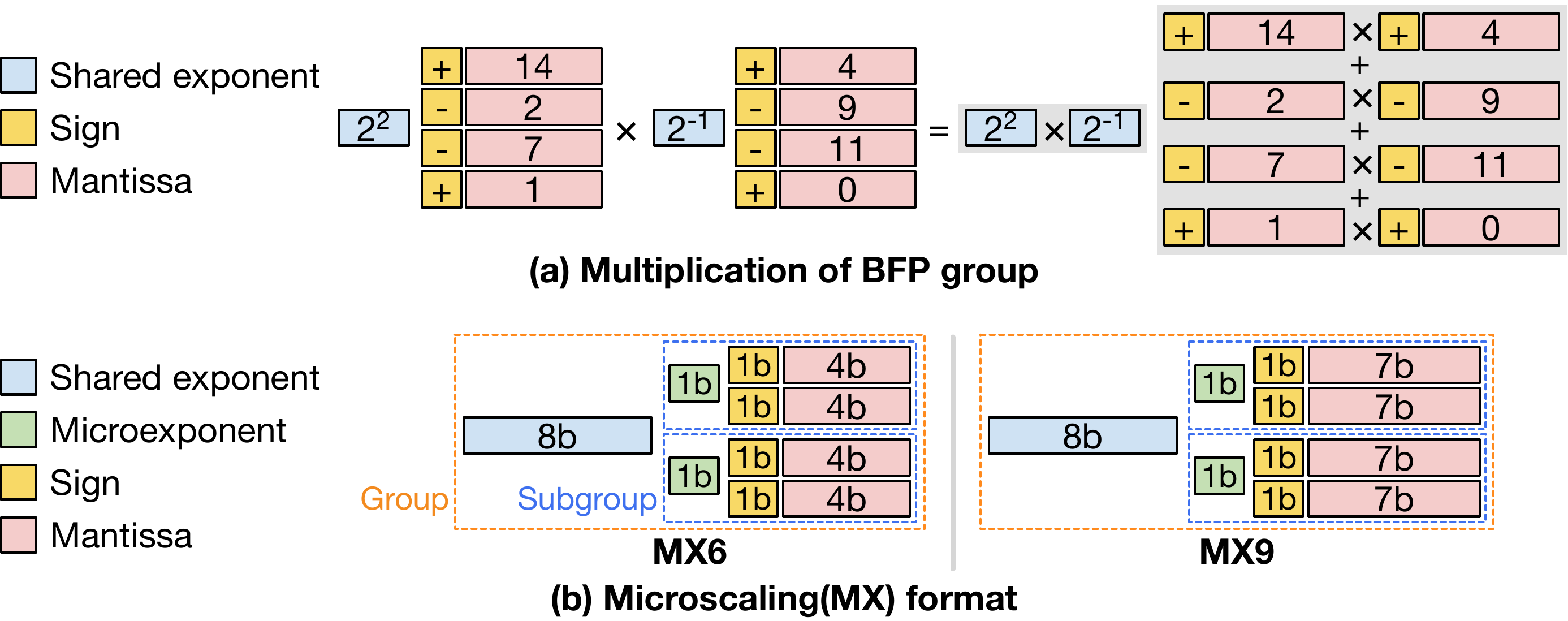}
\caption{Block floating point and microscaling (MX) formats. While our design uses a group size of 16, the illustration depicts a group size of 4 for clarity.}
\label{fig:bfp}
\end{figure}

\IEEEpubidadjcol
\niparagraph{Microscaling (MX) format.}
Block floating point (BFP) is one of the low-precision formats that has gained attention for balancing accuracy and computational efficiency.
Figure~\ref{fig:bfp}(a) shows how BFP multiplication operates.
Unlike conventional floating point, BFP simplifies multiplication by grouping values that share an exponent.
Microscaling (MX) format~\cite{mx, ocp} is a variant of BFP, as shown in Figure~\ref{fig:bfp}(b).
MX has demonstrated potential in LLMs but has not yet been explored for DiTs.
This work first characterizes the challenges of applying MX to DiT-based models and then leverages these insights to develop \emph{mixed-precision} MX quantization for DiTs.
%


\vspace{-0.3cm}
\subsection{Challenges of Applying MX formats to DiT}
\begin{figure}[t]
\centering
\includegraphics[width=0.8\linewidth]{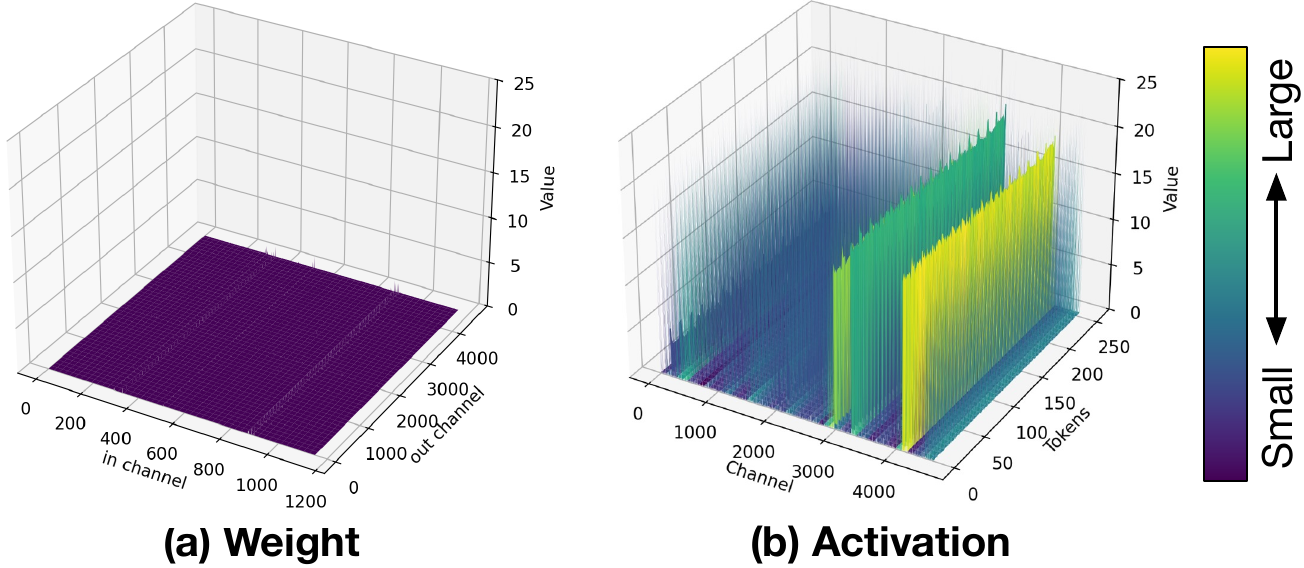}
\caption{Magnitude distributions of weights and activations in DiT-XL-256.}
\label{fig:distribution}
\end{figure}

\vspace{-0.15cm}
\niparagraph{Activation sensitivity to low precision.}
Table~\ref{tab:mx-degrade} presents the image quality (FID; lower is better) after applying the MX format to Stable Diffusion 3.
Applying MX6 to the weights while using MX9 for the activations preserves image quality comparable to FP16, whereas applying MX6 to both weights and activations causes a noticable degradation in image quality.
This is due to the value distribution of DiTs.
Figure~\ref{fig:distribution} illustrates the magnitude distribution of values in the weight and activation matrices, revealing more pronounced outliers in the activation matrix.
These outliers in activations significantly contribute to degradation.

\begin{figure}[t]
\centering
\includegraphics[width=0.75\linewidth]{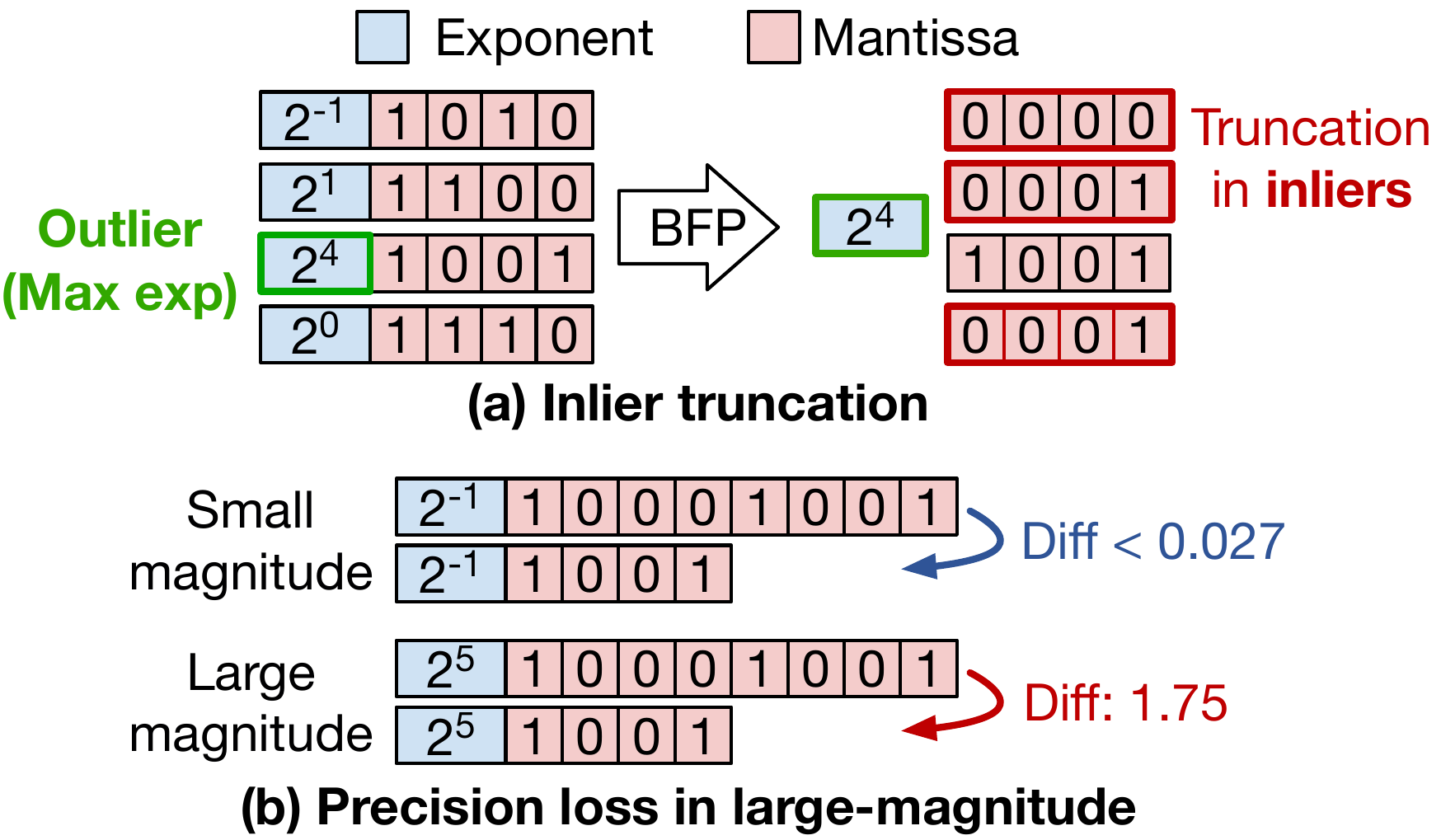}
\caption{Impact of large-magnitude values on MX quantization degradation.}
\label{fig:truncation}
\end{figure}

\niparagraph{Source of quality degradation.}
Figure~\ref{fig:truncation} illustrates two primary ways in which large-magnitude values degrade quality.
First, inliers within the same group as outliers are truncated because the shared group exponent is set to the largest exponent in the group, leading to precision loss in smaller values.
Second, large-magnitude values experience significant quantization error when represented with a 4-bit mantissa, whereas small-magnitude values remain more precise.
This work proposes a mixed-precision approach to mitigate these degradations caused by large-magnitude values.
%


\begin{table}[t]
\centering
\footnotesize
\caption{Image quality after applying MX to Stable Diffusion 3}
\label{tab:mx-degrade}
\begin{tabular}{ccc} 
\hline
Method              & Precision (W/A) & FID in COCO-1k ($\downarrow$)  \\ 
\hline
FP                  & 16/16           & 74.07                       \\
\hdashline[1pt/1pt]
\multirow{4}{*}{MX} & 9/9             & 72.98                       \\
                    & 6/9             & 71.88                       \\
                    & 9/6             & 203.37                      \\
                    & 6/6             & 199.78                      \\
\hline
\end{tabular}
\end{table}

\vspace{-0.3cm}
\section{Design}
\vspace{-0.1cm}
This section describes the MX-based mixed-precision quantization scheme and the \sysname accelerator architecture. 
%

\vspace{-0.35cm}
\subsection{Magnitude-based Mixed-Precision MX Quantization}
\begin{figure}[t]
\centering
\includegraphics[width=0.8\linewidth]{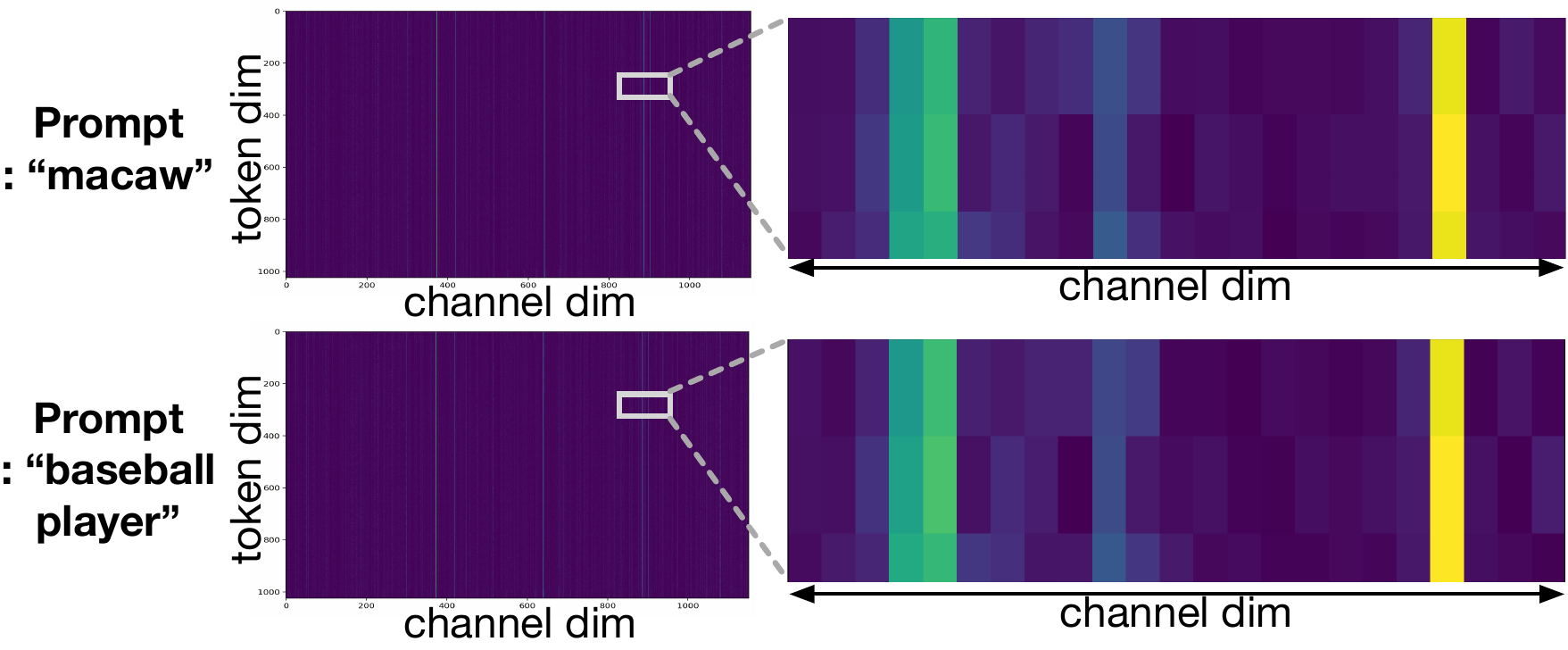}
\caption{Observation of linear layer activation's value magnitude in DiT-XL-512. The colors mean same with Fig.~\ref{fig:distribution}.}
\label{fig:observation}
\end{figure}

\begin{figure}[t]
\centering
\includegraphics[width=0.9\linewidth]{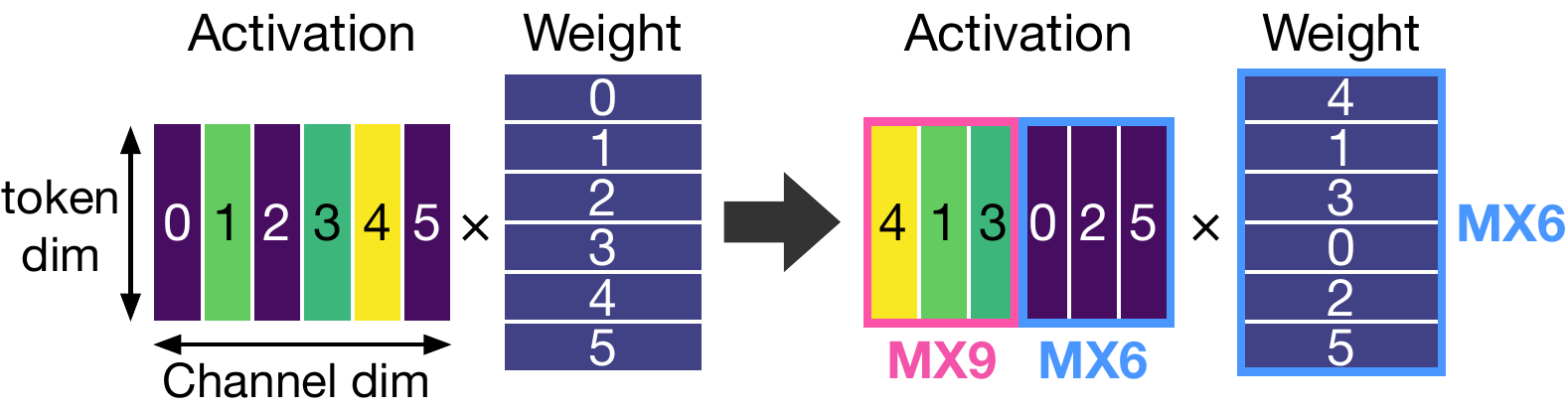}
\caption{Channel-wise reordering and mixed precision in linear layer. The colors mean same with Fig.~\ref{fig:distribution}.}
\label{fig:linear}
\end{figure}

\niparagraph{Mixed-precision quantization for linear layers.}
Figure~\ref{fig:observation} reports that large-magnitude values concentrate in specific channels, while outliers consistently appear in the same ones regardless of the prompt~\cite{qdit}. 
Given the observation, we propose a mixed-precision quantization scheme for linear layers, utilizing a channel-wise reordering technique as shown in Figure~\ref{fig:linear}.
\sysname first reorders channels according to their average magnitude, clustering outliers and inliers together to prevent the inlier truncation caused by outliers within the same group.
It then reorders weight channels following the same order as the activation channels.
After reordering, \sysname quantizes activation channels with large-magnitude values using high precision (MX9), while applying low precision (MX6) to channels with small-magnitude values and weight matrices.
The proportion of activation channels quantized with MX9 is controlled by a hyperparameter, $p_1$, whose its determination mechanism will be discussed in Section~\ref{subsec:hyperparam-selection}. 
%

\begin{figure}[t]
\centering
\includegraphics[width=0.9\linewidth]{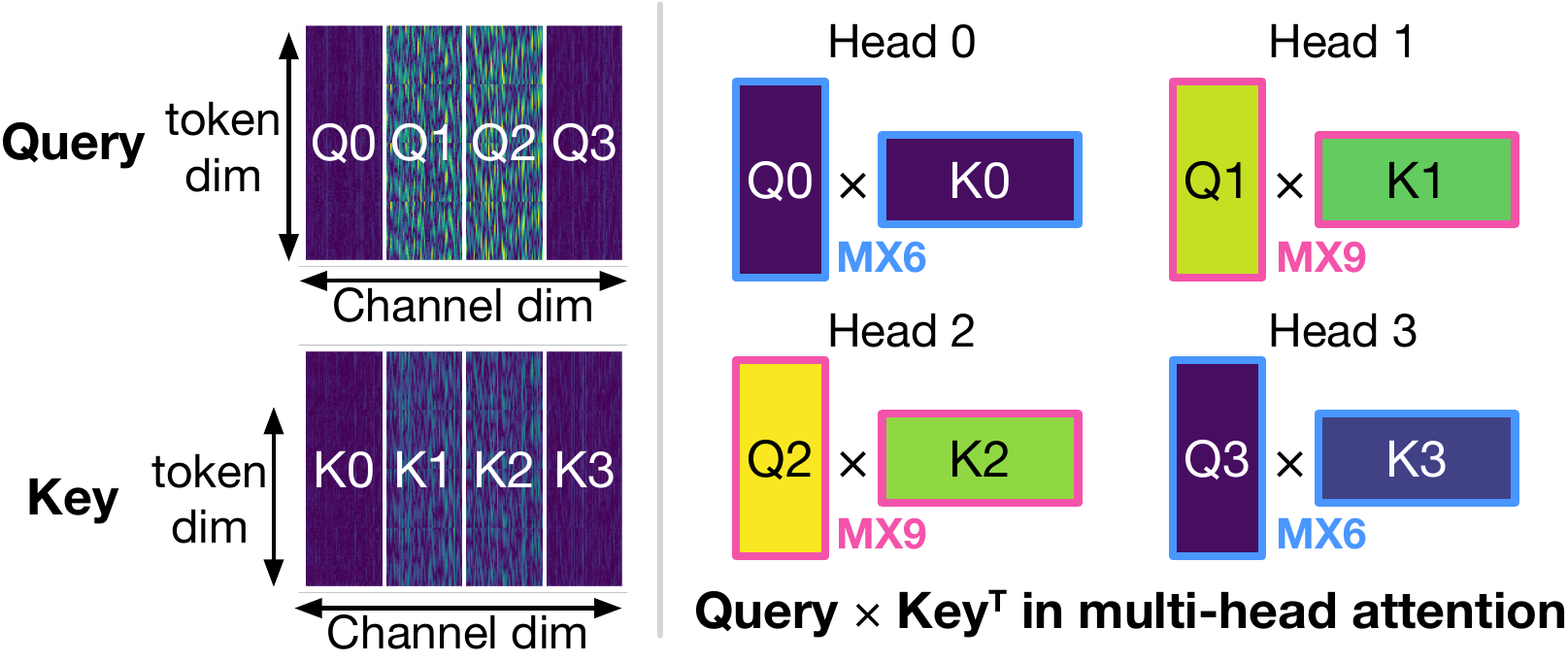}
\caption{Head-wise mixed precision in multi-head attention layer. The colors mean same with Fig.~\ref{fig:distribution}.}
\label{fig:attention}
\end{figure}

\niparagraph{Mixed-precision quantization for attention layers.}
The attention layer, performing activation-activation multiplication, requires a different approach.
Figure~\ref{fig:attention} delineates the scheme.
In multi-head attention, each head independently performs the $Q\times K^T$ and $Softmax\times V$ operations. 
Unlike the linear layer, where outliers are present at the channel level, the attention layer exhibits them at the head level. 
Analyzing 1,000 COCO prompts, we observed that as in channels in the linear layer, large-magnitude heads remain consistent across various inputs, allowing offline identification.
In our head-wise mixed-precision design, heads with large average magnitudes are quantized with MX9, while those with small magnitudes use MX6. 
As with the linear layer, the proportion of heads quantized with MX9 is controlled by a hyperparameter, $p_2$.
%
%

\vspace{-0.3cm}
\subsection{Offline Hyperparameter Determination} 
\begin{figure}[t]
\centering
\includegraphics[width=0.8\linewidth]{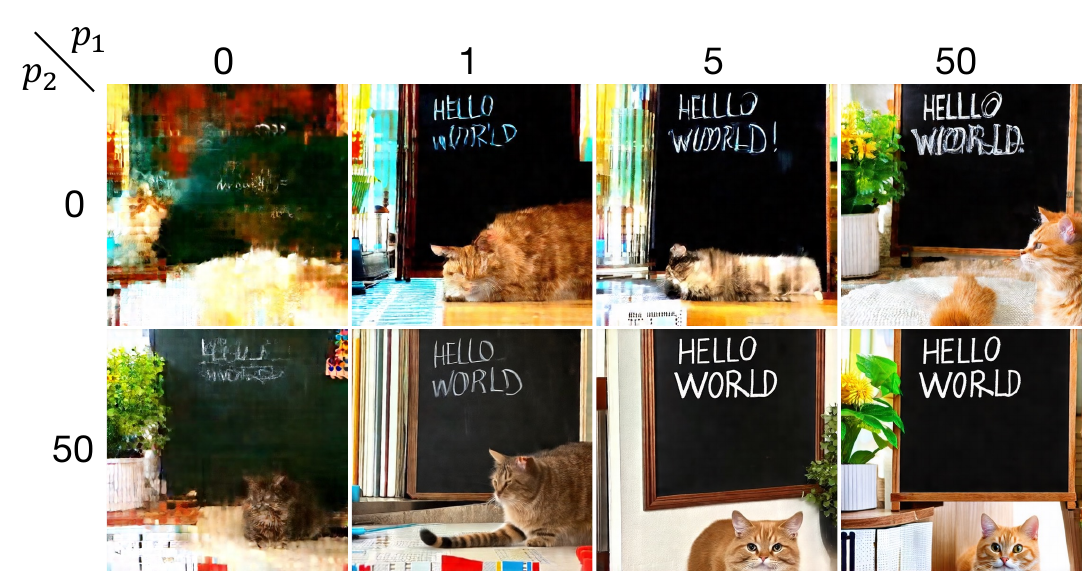}
\caption{Image quality-latency tradeoff according to the hyperparameter. Input prompt: ``hello world'' written on the blackboard and a cat.}
\label{fig:tradeoff}
\end{figure}

\vspace{-0.1cm}
\label{subsec:hyperparam-selection}
\niparagraph{Importance of the hyperparameters.}
Figure~\ref{fig:tradeoff} shows the quality-latency tradeoff according to the hyperparameter $p_1$ and $p_2$.
As the $p_1$ and $p_2$ increase, a larger portion of the values are quantized with high precision, leading to improved image quality. 
However, this comes at the cost of increased latency due to the high-precision computations. 
Additionally, the optimal values of $p_1$ and $p_2$ vary across models, while their dependence on data remains marginal.
%


\niparagraph{Hyperparameter determination algorithm.}
Inspired by these insights, we design an offline hyperparameter determination algorithm.
%
\sysname sweeps through possible $p_1$ and $p_2$ candidates, generating 64 images for each parameter configuration. 
Then, the algorithm determines $p_1$ and $p_2$ by jointly considering (1) the quality of the generated images (measured by FID), and (2) the latency in the hardware.
More specifically, we select ($p_1$, $p_2$) with minimum $FID\times latency^{\alpha}$. 
While we empirically set $\alpha$ to 0.15 by default, it can be adjusted to prioritize quality or latency.
A higher $\alpha$ emphasizes shorter latency, while a lower $\alpha$ prioritizes higher quality. 


\vspace{-0.4cm}
\subsection{\sysname Accelerator Architecture}
\begin{figure}[t]
\centering
\includegraphics[width=0.85\linewidth]{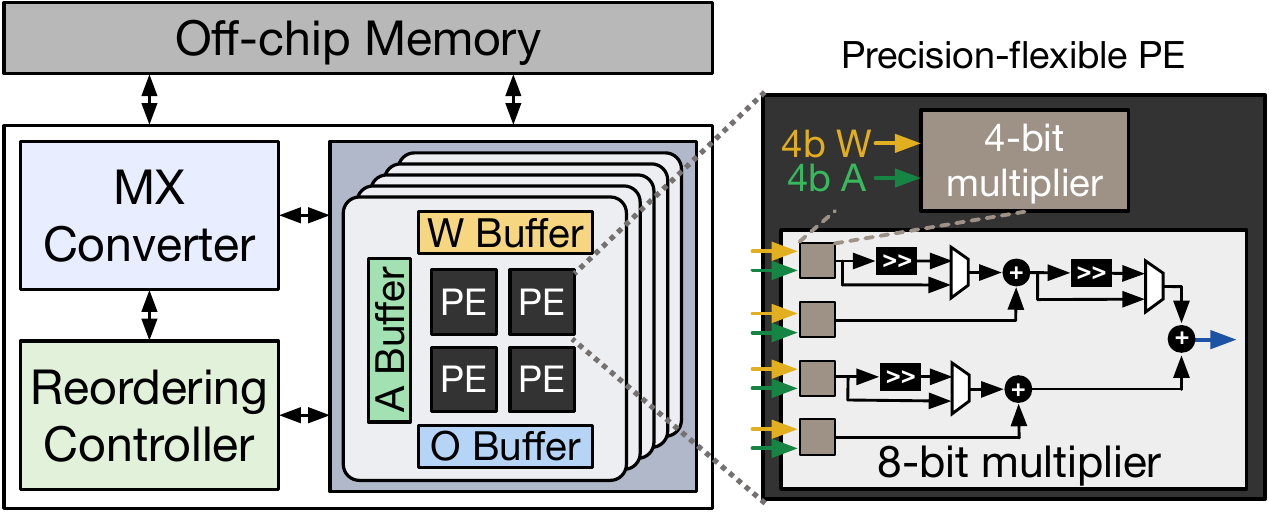}
\caption{\sysname accelerator architecture, built around a systolic array with an MX converter and reordering controller for precision-flexible MX processing.}
\label{fig:hardware}
\end{figure}

\vspace{-0.1cm}
\niparagraph{Architecture for precision-flexible MX quantization support.}
Figure~\ref{fig:hardware} presents an overview of \sysname architecture.
The accelerator is centered around systolic arrays, which are attached with a reordering controller and an MX converter.
Systolic arrays perform computations by fetching the input and weight matrices, pre-ordered in MX format, from off-chip memory.
After the computation, the reordering controller selects the appropriate output matrix channels from each bank of the output buffer. 
For the reordering, the controller maintains a table that specifies the required channel order for each layer and timestep.
It forwards these channels in the required order to MX converter.
The MX converter then groups and converts the reordered matrix into MX format before storing it in off-chip memory. 
The MX converter is composed of combinational logic, resulting in negligible latency.
The accelerator components operate in a pipelined manner, and the latencies of reordering controller and MX converter are effectively hidden by the significantly larger latency of the systolic array.

\niparagraph{MX systolic array.}
Our mixed precision technique handles MX6$\times$MX9 (for linear layers), MX9$\times$MX9 (for attention layers), and MX6$\times$MX6 operations. 
The processing elements in the systolic array should support these three types of multiplication. 
We adopted precision-flexible processing element for MX operations proposed in DaCapo~\cite{dacapo}. 
This PE has four 4-bit multipliers that can do the dot product of the 4-bit mantissa.
Therefore, when the group size is 16, MX6$\times$MX6, which needs 4-bit mantissa multiplication, takes 4 cycles per group dot product.
MX6$\times$MX9 and MX9$\times$MX9 need 8-bit mantissa multiplication.
The outputs of the four 4-bit multipliers mentioned above become one 8-bit multiplier output.
Therefore, it takes 16 cycles per group dot product.



\vspace{-0.4cm}
\section{Methodology}
\vspace{-0.1cm}

\begin{table}
\centering
\footnotesize
\caption{Hardware configurations}
\label{tab:hw-config}
\begin{tabular}{c;{1pt/1pt}c|c;{1pt/1pt}c} 
\hline
Systolic array size   & 16x16 PEs & Memory bandwidth        & 936GB/s   \\ 
\hdashline[1pt/1pt]
\# of systolic arrays & 1024      & On-chip memory & 28MB      \\ 
\hdashline[1pt/1pt]
Frequency             & 500 MHz   & Peak perf.(MX9)   & 262 TOPS  \\
\hline
\end{tabular}
\end{table}


\niparagraph{Models and datasets.}
For evaluation, we use DiT-XL~\cite{dit}, Pixart-$\Sigma$~\cite{pixart-sigma}, and Stable Diffusion 3 (SD3)~\cite{sd3}.
We denote the generated image resolution by appending $\{256, 512, 1024\}$ to the model name.
We use the default settings for the guidance scale of 4.0 and 5.0 for DiT-XL and Stable Diffusion 3, respectively.
We perform inference for 25 timesteps across all models.
To evaluate the quality, we employ ImageNet-val-5k for DiT-XL, and MS COCO-1k for Pixart-$\Sigma$ and SD3. 
We adopt three image quality metrics: fidelity with Frechet Inception Distance (FID), diversity with Inception Score (IS), and prompt alignment with CLIP Score. 

\niparagraph{Baselines.}
%
We employ Q-DiT~\cite{qdit} and ViDiT-Q~\cite{viditq} as quality baselines, both of which are prior works using INT quantization for diffusion transformers.
We set the precision to 6 bits for both weights and activations to demonstrate their limitations in low-precision activation quantization.
As speedup baselines, we use the original FP16 model and the ViDiT-Q W8A8 model on RTX 3090.
Since ViDiT-Q provides an INT8 model only for Pixart-$\Sigma$, we conduct our comparison with ViDiT-Q exclusively on Pixart-$\Sigma$.

\niparagraph{Implementation.}
We implement the MX quantized model with PyTorch 2.0, Huggingface Diffusers library, and triton language~\cite{triton-language}. 
We develop our accelerator on top of the open-source DaCapo accelerator simulator~\cite{dacapo}.
%
%
Table~\ref{tab:hw-config} shows the hardware configuration for the simulator.
We set the group size to 16 and the subgroup size to 2 for MX, and employ a batch size of 1 throughout the experiments.

\vspace{-0.3cm}
\section{Evaluation}
\begin{figure}[t]
\centering
\includegraphics[width=0.8\linewidth]{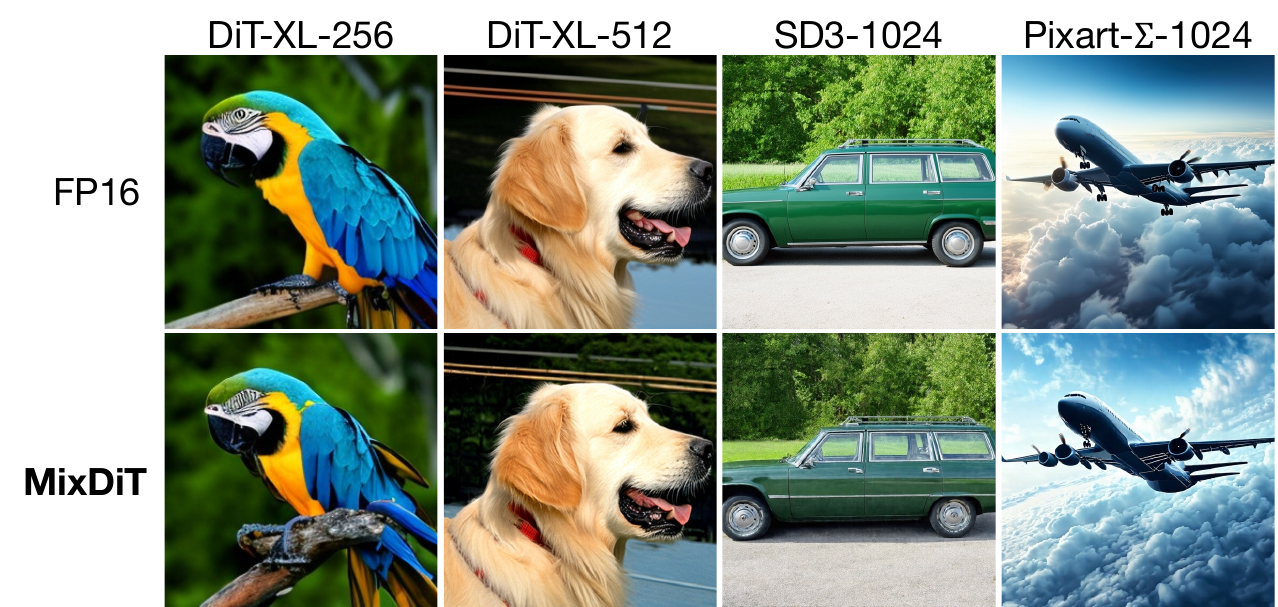}
\caption{Images generated by FP16 model and \sysname.}
\label{fig:quality}
\end{figure}

\vspace{-0.1cm}
\niparagraph{Image quality.}
Figure~\ref{fig:quality} visualizes the output image quality of the FP16 model and \sysname, demonstrating that \sysname's output image quality matches that of FP16. 
Table~\ref{tab:quality} reports quantitative results for image quality.
%
%
As Q-DiT and ViDiT-Q are designed to use high activation bitwidths (e.g., INT8/8 or INT4/8), they suffer from poor image quality when using INT6/6 precision.
In contrast, our mixed-precision scheme applies MX9 only to activation outliers while using MX6 for the rest, achieving image quality comparable to FP16.
%


\begin{table}
\caption{Image quality evaluation}
\label{tab:quality}
\centering
\footnotesize
\begin{tabular}{cccccc} 
\toprule
\begin{tabular}[c]{@{}c@{}}Model\\($p_1$, $p_2$)\end{tabular}                        & \begin{tabular}[c]{@{}c@{}}Precision\\(W/A)\end{tabular} & Method  & FID$\downarrow$   & IS$\uparrow$     & \begin{tabular}[c]{@{}c@{}}CLIP\\score$\uparrow$\end{tabular}  \\ 
\midrule
\multirow{4}{*}{\begin{tabular}[c]{@{}c@{}}DiT-XL-256\\(0, 0)\end{tabular}}    & FP (16/16)      & FP16    & 17.32 & 231.20 &  - \\
    & INT (6/6)       & Q-DiT       & 29.27             & 169.49            &  - \\
    & MX (6/6)        & MX6         & 67.27             & 59.99             &  - \\
    & MX (6/6)        & \sysname    & \textbf{15.39}    & \textbf{232.85}   & - \\ 
\midrule
\multirow{4}{*}{\begin{tabular}[c]{@{}c@{}}DiT-XL-512\\(1, 0)\end{tabular}}    & FP (16/16)      & FP16    & 20.55 & 216.83 & - \\
    & INT (6/6)       & Q-DiT       & \textbf{18.04}    & 209.77            & - \\
    & MX (6/6)        & MX6         & 108.70            & 22.88             & - \\
    & MX (6/6)        & \sysname    & 20.15             & \textbf{217.77}   & - \\ 
\midrule
\multirow{3}{*}{\begin{tabular}[c]{@{}c@{}}SD3-1024\\(5, 20)\end{tabular}}     & FP (16/16)      & FP16    & 74.07    &    -    & \textbf{29.56} \\
    & MX (6/6)        & MX6         & 199.78            &   -    & 22.74 \\
    & MX (6/6)        & \sysname    & \textbf{72.48}    &   -    & 29.34 \\ 
\midrule
\multirow{4}{*}{\begin{tabular}[c]{@{}c@{}}Pixart-$\Sigma$-1024\\(1, 20)\end{tabular}} & FP (16/16)      & FP16    & 69.96    &   -    & 31.34  \\
    & INT (6/6)       & ViDiT-Q     & 84.74             &   -    & \textbf{31.93} \\
    & MX (6/6)        & MX6         & 71.66             &   -    & 31.31          \\
    & MX (6/6)        & \sysname    & \textbf{69.29}    &   -    & 31.50            \\
\bottomrule
\end{tabular}
\end{table}

\begin{figure}[t]
\centering
\includegraphics[width=\linewidth]{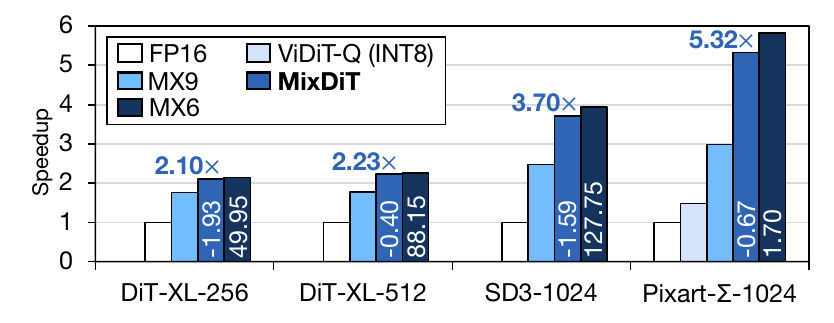}
\caption{Latency speedup compared to RTX 3090. The y-axis represents speedup (higher is better), while the numbers inside the bars indicate FID degradation compared to FP16 (lower is better).}
\label{fig:speedup}
\end{figure}

\niparagraph{Speedup.}
Figure~\ref{fig:speedup} reports \sysname latency speedup compared to the baseline schemes on NVIDIA RTX 3090.
%
When using MX9, our baseline accelerator achieves the same TOPS as the RTX 3090 with INT8.
%
Leveraging the mixed-precision of MX6 and MX9, \sysname achieves 2.10$\times$ to 5.32$\times$ speedup compared to the GPU. 
We observe that only $\leq$5\% of activations in linear layers and $\leq$20\% of activations in attention layers are quantized with MX9, allowing \sysname to achieve similar speedups to MX6 while minimizing quality loss.
%
Quantizing most data to MX6 leverages the 4-bit multipliers, which accelerate systolic array computations and serve as the main contributor to the speedup.
We notice that larger image generation models see greater improvement, as GEMM computations dominate latency.

\vspace{-0.3cm}
\section{Conclusion}
\vspace{-0.1cm}
This paper presents \sysname, a mixed-precision quantization framework for accelerating image diffusion transformers using MX formats. 
Our approach applies higher precision to activation outliers while using low-precision MX computations for the rest, maintaining high image quality. 
To fully exploit mixed precision, we design a specialized hardware architecture for efficient MX quantization.
Combining algorithmic and architectural innovations, \sysname enables low-precision inference while preserving generation quality, offering a practical solution for efficient diffusion transformer acceleration. 
%


\vspace{-0.3cm}
\bibliographystyle{IEEEtran}

\vfill

\end{document}